\documentclass[11pt,a4paper]{article}
\usepackage{jcappub,epsfig,multicol,bbm,latexsym,graphicx,subfigure}

\def\prl{Phys.\ Rev.\ Lett.\ }
\def\aap{Astron.\ Astrophys.\ }

\def\apj{Astrophys.\ J.\ }
\def\apjl{Astrophys.\ J.\ Lett.\ }
\def\apjs{Astrophys.\ J.\ Suppl.\ }

\def\mnras{Mon.\ Not.\ Roy.\ Astron.\ Soc.\ }
\def\physrep{Phys.\ Rept.\ }
\def\prd{Phys.\ Rev.\ D\ }

\def\jcap{JCAP}

\def\mchi{\ensuremath{m_\chi}}
\def\sv{\ensuremath{\langle\sigma v\rangle}}

\title{Revisit of cosmic ray antiprotons from dark matter annihilation with 
updated constraints on the background model from AMS-02 and collider data}

\author{Ming-Yang Cui$^{a,b}$, Xu Pan$^a$, 
Qiang Yuan$^{b,c,d}$\footnote{Corresponding author.}, 
Yi-Zhong Fan$^{b,c}$ and Hong-Shi Zong$^a$\footnote{Corresponding author.}}

\affiliation{
$^a$School of Physics, Nanjing University, Nanjing 210093, China \\
$^b$Key Laboratory of Dark Matter and Space Astronomy, Purple
Mountain Observatory, Chinese Academy of Sciences, Nanjing
210008, China \\
$^c$School of Astronomy and Space Science, University of Science and
Technology of China, Hefei, Anhui 230026, China \\
$^d$Center for High Energy Physics, Peking University, Beijing 100871, China}
\emailAdd{mycui@pmo.ac.cn}
\emailAdd{xupan@pmo.ac.cn}
\emailAdd{yuanq@pmo.ac.cn}
\emailAdd{yzfan@pmo.ac.cn}
\emailAdd{zonghs@nju.edu.cn}

\abstract{We study the cosmic ray antiprotons with updated constraints
on the propagation, proton injection, and solar modulation parameters
based on the newest AMS-02 data near the Earth and Voyager data in the 
local interstellar space, and on the cross section of antiproton 
production due to proton-proton collisions based on new collider data.
We use a Bayesian approach to properly consider the uncertainties of
the model predictions of both the background and the dark matter (DM)
annihilation components of antiprotons. We find that including an extra 
component of antiprotons from the annihilation of DM particles into a 
pair of quarks can improve the fit to the AMS-02 antiproton data 
considerably. The favored mass of DM particles is about $60\sim100$ GeV, 
and the annihilation cross section is just at the level of the thermal 
production of DM ($\sv \sim O(10^{-26})$ cm$^3$~s$^{-1}$).}

\keywords{dark matter, cosmic ray}

\arxivnumber{1803.02163}

\begin{document}
\maketitle
\flushbottom

\section{Introduction}

Cosmic ray (CR) antiprotons are one of the most important probes to
indirectly detect dark matter (DM) particles. Quite a number of
balloon and space detectors have been dedicated to precisely measuring 
the antiproton fluxes and antiproton-to-proton ratios since 1990s
\cite{1996PhRvL..76.3057M,1997ApJ...487..415B,1999ICRC....3...77B,
2000PhRvL..84.1078O,2001APh....16..121M,2001ApJ...561..787B,
2001PhRvL..87A1101B,2002PhR...366..331A,2002PhRvL..88e1101A,
2005ICRC....3...13H,2008PhLB..670..103A,2009PhRvL.102e1101A,
2010PhRvL.105l1101A,2016PhRvL.117i1103A}. Large progresses have
been made in recent years thanks to the operation of the Payload 
for Antimatter Matter Exploration and Light-nuclei Astrophysics
(PAMELA; \cite{2018arXiv180110310A}) and the Alpha Magnetic 
Spectrometer (AMS-02). The antiproton flux has been measured to
$\sim500$ GeV by AMS-02 with high precision~\cite{2016PhRvL.117i1103A}, 
which improved the constraints on either the CR or the DM models 
effectively~\cite{2016PhRvD..94l3019K,2016ApJ...824...16J,
2015PhRvD..92e5027J,2017PhRvD..96l3010L,2016PTEP.2016b1E01K,
2017PhRvD..96b3006L,2017PhRvD..95l3007C,2017PhRvD..95f3021H,
2017JHEP...04..112L,2017arXiv170102263F}.

The propagation of CRs is typically one of the largest sources of 
uncertainties in predicting the background of antiprotons and the signal 
from DM annihilation~\cite{2001ApJ...563..172D,2004PhRvD..69f3501D,
2015JCAP...03..021H}. The propagation of charged particles in the Milky 
Way can be described by a diffusion process in the random magnetic field. 
The collision between primary nuclei and the interstellar gas leads to 
fragmentation of the parent nuclei and the production of secondary nuclei. 
Therefore the secondary-to-primary particle ratio, e.g., the Boron-to-Carbon
(B/C) ratio, is usually employed to constrain the propagation 
parameters~\cite{2001ApJ...555..585M,2011ApJ...729..106T,
2015JCAP...09..049J,2016PhRvD..94l3007F}. The precise measurements of 
CR fluxes and/or flux ratios by AMS-02~\cite{2016PhRvL.117i1103A,
2016PhRvL.117w1102A} shed new light on the understanding of CR 
propagation~\cite{2016PhRvD..94l3019K,2016ApJ...824...16J,
2017PhRvD..95h3007Y,2018PhRvD..97b3015N,2018JCAP...01..055R}.

Using the improved constraints on the propagation model parameters
\cite{2016PhRvD..94l3019K,2017PhRvD..95h3007Y}, it was found that 
there might be an excess of the antiproton flux around a few GeV 
energies compared with the background contribution from $pp$ collisions, 
and a DM model could simply explain this excess without constraints 
from other observations such as $\gamma$-rays~\cite{2017PhRvL.118s1101C,
2017PhRvL.118s1102C}. More interestingly, the model parameters to
account for the antiproton excess are consistent with that proposed to 
explain the Galactic center $\gamma$-ray excess~\cite{2011PhLB..697..412H,
2012PhRvD..86h3511A,2015PhRvD..91l3010Z,2016JCAP...04..030H},
the tentative $\gamma$-ray excesses in the directions of two dwarf
galaxies~\cite{2015PhRvL.115h1101G,2016PhRvD..93d3518L}, and a possible
$\gamma$-ray line-like feature from a population of clusters of galaxies 
\cite{2016PhRvD..93j3525L}. Such coincidence makes DM a promising
explanation of the possible antiproton excess~\cite{2017JCAP...10..053C}.

Most recently, the AMS-02 collaboration reported new measurements 
of the primary (He, C, O) and secondary (Li, Be, B) nucleus fluxes
\cite{2017PhRvL.119y1101A,2018PhRvL.120b1101A}. These results are
expected to give more consistent constraints on the CR propagation models 
and parameters since they are closely relevant parent and daughter 
particles (different from the B/C ratio and proton fluxes used in 
Ref.~\cite{2017PhRvD..95h3007Y}). Using the Carbon flux and B/C ratio 
measured by AMS-02~\cite{2016PhRvL.117w1102A,2017PhRvL.119y1101A},
together with the data at low energies by the Advanced Composition 
Explorer (ACE) spacecraft near the Earth and the Voyager in the local 
interstellar space~\cite{2013Sci...341..150S,2016ApJ...831...18C}, 
Ref.~\cite{2018arXiv180510649Y} carried out a study of different 
propagation model settings and constrained the propagation parameters 
in a narrow region. In this work, we revisit the antiproton
problem based on the new results of the CR propagation. The updated 
production cross section of antiprotons via $pp$ collisions with
constraints from the most recent collider data will also be employed
\cite{2017JCAP...02..048W}. In addition, we also develop a more efficient 
method to calculate the likelihood of the DM component. In Section 2 we 
present the propagation and proton source parameters which are the basis 
of the calculation of the background antiproton flux. In Section 3 we 
investigate the DM contribution to antiprotons. We conclude our work 
in Section 4.

\section{Propagation model parameters and background antiprotons}

Charged particles propagate diffusively in a diffusive halo, usually
assumed to feature cylindrical symmetry, with a radius of $R_h$ and a 
half-height $z_h$, defined by the extension of the magnetic field. 
The propagation equation can be written in general as 
\begin{equation}
\frac{\partial \psi}{\partial t}=\nabla\cdot(D_{xx}\nabla\psi-{\bf V}\psi)
+\frac{\partial}{\partial p}p^2D_{pp}\frac{\partial}{\partial p}
\frac{\psi}{p^2}+\frac{\partial}{\partial p}\left[b\psi+\frac{p}{3}
(\nabla\cdot{\bf V})\psi\right]-\frac{\psi}{\tau}+q.
\label{eq:prop}
\end{equation}
The first term in the right-hand-side is the diffusion in the random
magnetic field with $D_{xx}$ being the spatial diffusion coefficient,
the second term represents the advection velocity which is assumed to 
linearly increase along the $z$-direction,
the third term is the stochastic reacceleration characterized by a
diffusion in the momentum space with a diffusion coefficient $D_{pp}$,
the fourth and fifth terms are the interaction and adiabatic energy 
loss terms, the sixth term represents the fragmentation and/or decay, 
and the last term is the source function. 

The diffusion coefficient is usually assumed to be spatially uniform in 
the Milky Way, and it can be parameterized as a power-law function of 
rigidity, $D(R)=\beta D_0(R/R_0)^{\delta}$, where $\beta$ is the 
velocity of a particle (in unit of light speed), $D_0$ is a normalization 
constant, $\delta$ is the rigidity-dependence slope (see below for a
modification of the diffusion coefficient). There are proposals of 
spatially non-uniform (e.g., \cite{2016PhRvD..94l3007F,2016ApJ...819...54G}) 
or anisotropic diffusion \cite{2017JCAP...10..019C} scenarios motivated by 
recent observations of spectral hardenings of CRs
\cite{2010ApJ...714L..89A,2011Sci...332...69A} and spatial variation 
of CR spectral indices inferred from Fermi-LAT $\gamma$-ray 
observations~\cite{2016PhRvD..93l3007Y,2016ApJS..223...26A}. 
It has been shown that the $\bar{p}/p$ ratio also hardens gradually 
at high energies in the spatially-dependent propagation model
\cite{2016PhRvD..94l3007F,2016ApJ...819...54G}, and may account for 
the flat behavior of the $\bar{p}/p$ ratio as observed by
AMS-02~\cite{2016PhRvL.117i1103A}. The effect on the low energy part
of the antiproton spectrum (e.g., below 10 GeV) under such complicated
propagation models needs further studies. Here we work under the simple
uniform diffusion framework, which can actually explain most of the
CR and diffuse $\gamma$-ray data.

The advection velocity is parameterized as ${\bf V}=dV/dz\cdot{\bf z}$.
The reacceleration is characterized by the Alfvenic speed of the plasma,
$v_A$, which bridges the spatial and momentum diffusion coefficients as 
$D_{pp}D_{xx}=\frac{4p^2v_A^2}{3\delta(4-\delta^2)(4-\delta)}$
\cite{1994ApJ...431..705S}. The momentum loss rate $b(p)=-\dot{p}$
includes the ionization and Coulomb scattering losses and radiative 
cooling (for electrons/positrons). The injection spectrum of nuclei is 
parameterized as a doubly broken power-law form of rigidity
\begin{equation}
q(R)\propto\left\{
\begin{array}{ll}
\left( R / R_{\mathrm{br},1}\right)^{-\nu_1}, & R < R_{\mathrm{br,1}} \\
\left( R / R_{\mathrm{br},1} \right)^{-\nu_2}, & R_{\mathrm{br,1}} \le R < R_{\mathrm{br,2}} \\
\left( R / R_{\mathrm{br},1} \right)^{-\nu_3}\left( R_{\mathrm{br},2} / R_{\mathrm{br},1} \right)^{\nu_3-\nu_2}, & R_{\mathrm{br,2}} \le R 
\end{array},\right.
\end{equation}
where $R_{\mathrm{br},1}\sim$ GV is to account for the low energy data,
and $R_{\mathrm{br},2}\sim300$ GV is to account for the high energy
spectral hardening~\cite{2010ApJ...714L..89A,2011Sci...332...69A}.

We assume the force-field approximation to describe the solar modulation 
effect~\cite{1968ApJ...154.1011G}. Since the time periods of data taking 
of protons and antiprotons by AMS-02 are slightly different, their
modulation parameters should also be different. In this work we adopt 
$\Phi_{\bar{p}}=\Phi_p+0.02$ GV, as suggested by the time-dependent solar 
modulation potentials \cite{2017PhRvD..95h3007Y}.

The numerical tool GALPROP~\cite{1998ApJ...509..212S,1998ApJ...493..694M} 
is adopted to solve the propagation equation.
It was found that the diffusion model with reacceleration of CRs by 
the random magnetohydrodynamic (MHD) waves fit the data better than the 
plain diffusion scenario and the model with a advective transportation
\cite{2018arXiv180510649Y}. In particular, a variant of the 
velocity-dependence of the diffusion coefficient, 
$D(R)=\beta^{\eta}D_0(R/R_0)^{\delta}$, where $\eta$ is an empirical 
modification of the velocity-dependence~\cite{2010APh....34..274D}, 
gives the best fit\footnote{The $\chi^2$ value of the DR2 model 
is 105.3 for 160 degrees of freedom. As a comparison, the $\chi^2$ 
values are 578.2 for the PD, 262.5 for the DC, 188.4 for the DC2, 
252.2 for the DR, and 248.6 for the DRC models, respectively. 
See Ref.~\cite{2017PhRvD..95h3007Y} for the definition of different 
propagation model settings.} to the data \cite{2018arXiv180510649Y}. 
Physically such a diffusion behavior may be related to the 
resonant interactions between CRs and the MHD waves which result in 
dissipations of such waves corresponding to low energy particles
\cite{2006ApJ...642..902P}. This model, referred to as DR2 hereafter, 
is employed to study antiprotons in this work. 

\begin{figure}[!htb]
\centering
\includegraphics[width=0.48\textwidth]{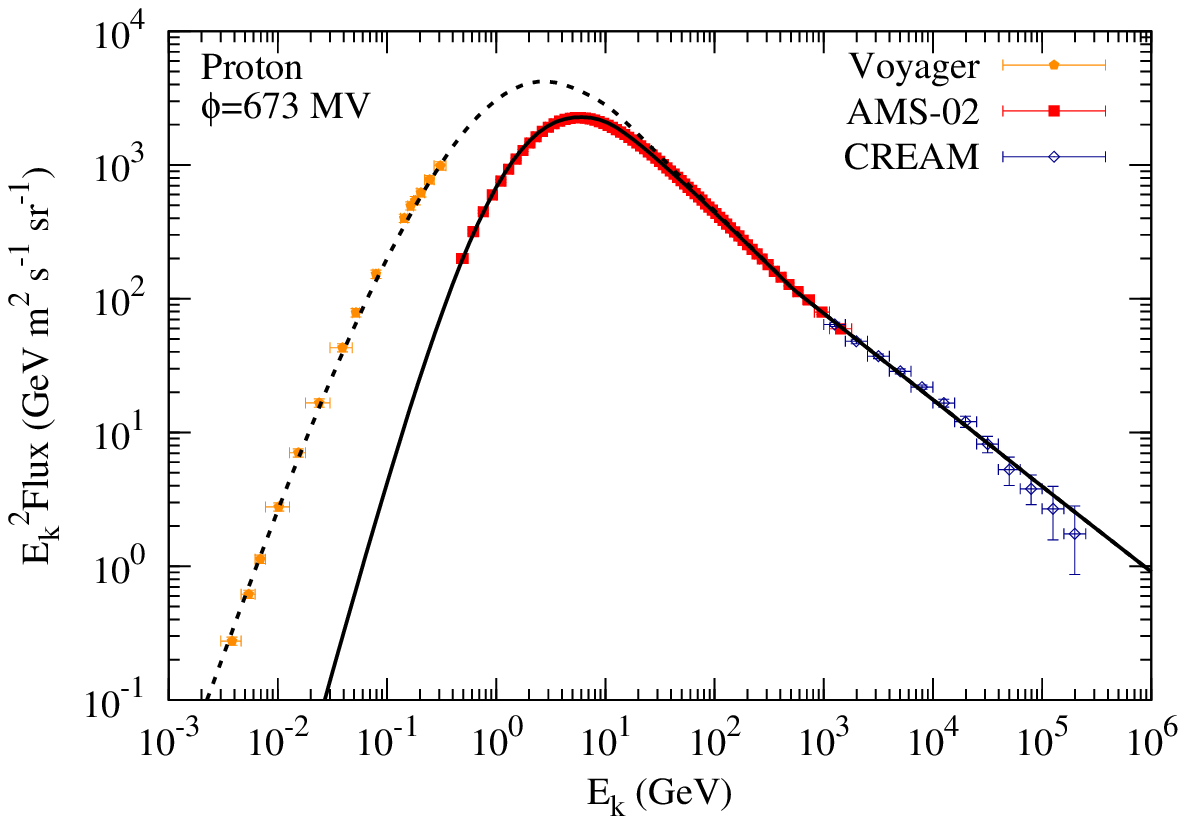}
\includegraphics[width=0.48\textwidth]{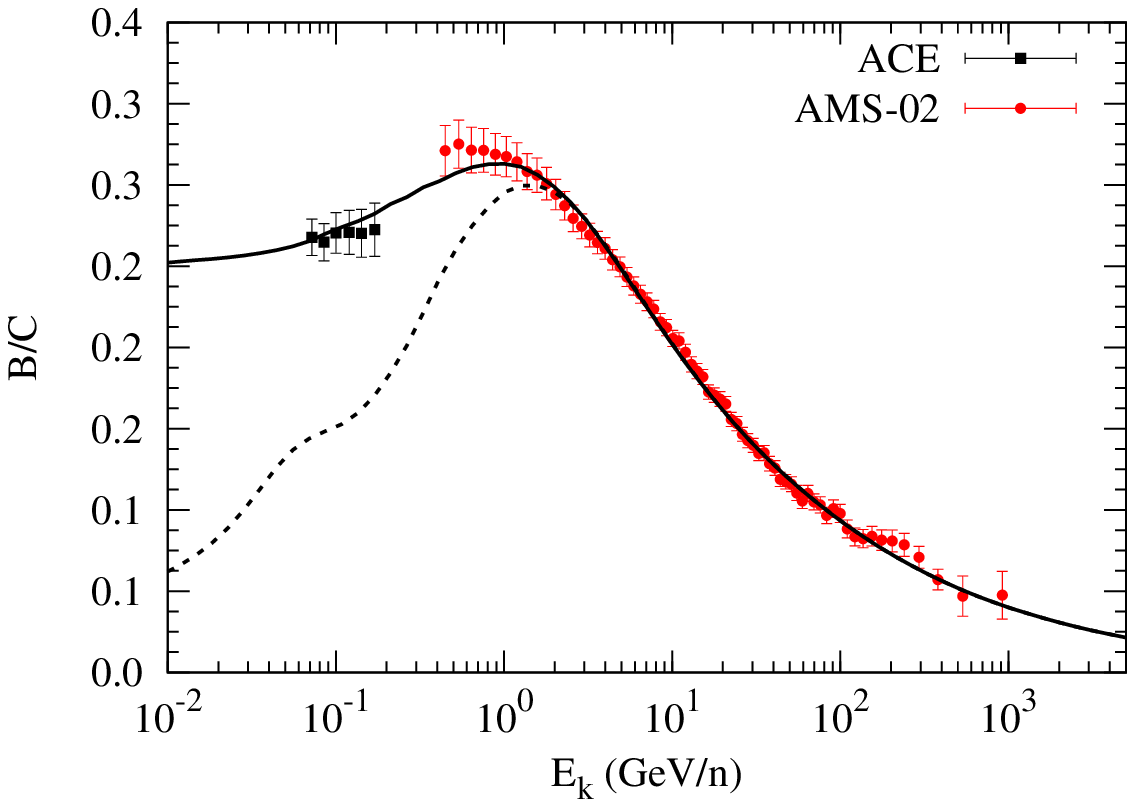}
\caption{Best-fit proton flux (left) and B/C ratio (right) compared with 
the Voyager, AMS-02, and CREAM data. In each panel the dashed (solid) 
line is for the result before (after) the solar modulation.
\label{fig:proton}}
\end{figure}

We use the {\tt CosRayMC} tool, which embeds the {\tt GALPROP} code 
in the Markov Chain Monte Carlo driver, to fit the model parameters 
\cite{2012PhRvD..85d3507L}. The proton fluxes measured by Voyager in 
the local interstellar space \cite{2016ApJ...831...18C}, 
AMS-02~\cite{2015PhRvL.114q1103A}, and CREAM~\cite{2017ApJ...839....5Y} 
are used to derive the proton 
injection spectral parameters. We include the uncertainties of the 
propagation parameters from independent fit to the B/C ratio and Carbon 
fluxes as priors. See the Appendix for details of the prior information
from the fitting covariance matrix. The fitting results of the source
parameters are given in Table~\ref{table:proton}. Some of the 
propagation parameters, such as $D_0$ and $z_h$, are consistent with
those derived previously in Ref.~\cite{2017PhRvD..95h3007Y}, while the
others are slightly different. This is perhaps due to different data 
sets used in this work (in particular the inclusion of Voyager data). 
The propagation parameters can not be directly compared with that given 
in Ref.~\cite{2017PhRvL.118s1102C} due to different model settings. 
However, we find that the parameters $D_0$ and $z_h$ are still similar
with each other. The best-fit results of the proton flux and the B/C 
ratio, together with the observational data, are shown in 
Figure~\ref{fig:proton}. Note that the injection spectrum of Carbon 
nuclei is different from that of protons, and we use the results obtained 
in Ref.~\cite{2018arXiv180510649Y} to calculate the B/C ratio. The 
propagation parameters used to plot Figure~\ref{fig:proton} are the same 
as that in Table~\ref{table:proton}.

\begin{table}[!htb]
\centering
\caption{Propagation and proton injection spectral parameters, 
together with their posterior $68\%$ errors, from fitting to the 
Voyager, AMS-02, and CREAM data.}
\begin{tabular}{ccc}
\hline \hline
Parameter & Unit & Value\\
\hline
$D_0^\dagger$ & ($10^{28}$cm$^2$ s$^{-1}$) & $5.98\pm1.00$ \\
$\delta$ & & $0.411\pm0.008$ \\
$z_h$ & (kpc) & $5.58\pm1.39$ \\
$v_A$ & (km s$^{-1}$) & $27.5\pm1.3$ \\
$\eta$ & & $-0.27\pm0.08$ \\
$A_p^\ddagger$ & ($10^{-9}$~cm$^{-2}$~s$^{-1}$~sr$^{-1}$) & $4.43\pm0.01$ \\
$\nu_1$ & & $2.03\pm0.02$ \\
$R_{\rm br,1}$ & (GV) & $10.3\pm0.9$ \\
$\nu_2$ & & $2.40\pm0.01$ \\
$R_{\rm br,2}$ & (GV) & $511.8\pm86.2$ \\
$\nu_3$ & & $2.25\pm0.02$ \\
$\Phi_p$ & (GV) & $0.673\pm0.015$ \\
\hline
\hline
\end{tabular}\\
$^\dagger$Normalization at $R_0=4$ GV.
$^\ddagger$Normalization of the propagated proton flux at 100 GeV.
\label{table:proton}
\end{table}

The antiproton background produced by inelastic collisions between
protons and the interstellar medium can be calculated using the same
propagation, proton injection, and solar modulation parameters
(referred to as background parameters hereafter) obtained through
fitting to the proton flux data. The Markov chains of the background
parameters are used, which include the correlations among different
parameters. The cross section of antiproton production is an
additional source of uncertainties \cite{1983JPhG....9.1289T,
2003PhRvD..68i4017D,2014PhRvD..90h5017D,2014JCAP...09..051K,
2015ApJ...803...54K,2017JCAP...02..048W,2018arXiv180203030K}.
In this work we employ the updated parameterization of the antiproton
production cross section based on the most recent collider data
\cite{2017JCAP...02..048W}. The relative uncertainties of the
antiproton fluxes are found to be $\lesssim 10\%$ in the relevant 
energy range covered by the AMS-02 data. Therefore we multiply a
constant factor $\kappa$, which has a Gaussian prior of $N(1.0,0.1^2)$
on the background antiproton flux when calculating its likelihood.

\section{DM contribution to antiprotons}

The same propagation parameters as obtained in Sec. II are adopted to
calculate the antiproton flux from the DM annihilation. The source
term of DM annihilation induced antiprotons can be written as
\begin{equation}
q_{\bar{p}}^{\rm DM}=\frac{\sv}{2\mchi^2}\frac{{\rm d}N}{{\rm d}E}
\rho(\boldsymbol{x}),
\end{equation}
where the DM particle is assumed to be Majorana fermion, $\mchi$
is the mass of the DM particle, $\sv$ is the velocity-weighted
annihilation cross section, ${\rm d}N/{\rm d}E$ is the antiproton
production spectrum per annihilation, and $\rho(\boldsymbol{x})$ is
the density profile of DM which is assumed to be the Navarro-Frenk-White 
distribution~\cite{1997ApJ...490..493N}. The scale radius of the density
profile is adopted to be 20 kpc, and the local density is assumed to 
be 0.3 GeV cm$^{-3}$ \cite{2016MNRAS.463.2623H}.

\begin{figure}[!htb]
\centering
\includegraphics[width=0.7\textwidth]{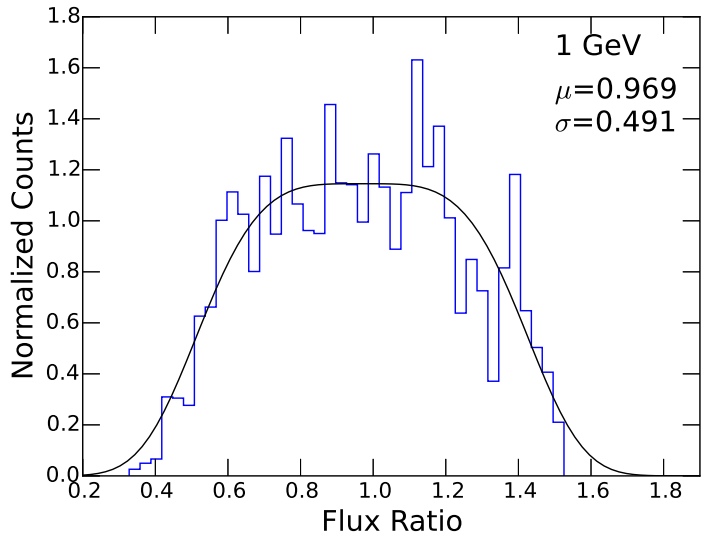}
\caption{Distribution of ratios of the DM-induced antiproton fluxes for
various background parameters $\boldsymbol{\theta}_{\rm bkg}$ to that 
calculated with the mean background parameters 
$\bar{\boldsymbol{\theta}}_{\rm bkg}$. The solid line is a fit using
Eq. (\ref{eq:probf}).
\label{fig:ratio}}
\end{figure}

We adopt a simplified way to calculate the likelihood of the DM 
component taking into account the uncertainties of the background 
parameters. The antiproton fluxes from the DM annihilation span in a 
band when the background parameters change \cite{2017PhRvL.118s1101C}. 
Figure~\ref{fig:ratio} shows the distribution of the ratios of antiproton 
fluxes at 1 GeV to that calculated with the mean background parameters 
given in Table~\ref{table:proton}. This distribution can be fitted with
a probability function
\begin{equation}
p(f)\propto \exp\left[-\left(\frac{f-\mu}{\sigma}\right)^4\right],
\label{eq:probf}
\end{equation}
where $\mu=0.969$ and $\sigma=0.491$. The posterior probability of a 
given set of DM parameters, $(\mchi,\,\sv)$, and specified annihilation
channel(s), is
\begin{equation}
\mathcal{P}_{\rm DM}\propto \int \mathcal{L}_{\bar{p}}
(\boldsymbol{\theta}_{\rm bkg},\kappa,f\bar{\phi}_{\rm DM})\,
p(\boldsymbol{\theta}_{\rm bkg})\,p(\kappa)\,p(f)\,
{\rm d}\boldsymbol{\theta}_{\rm bkg}\,{\rm d}\kappa\,{\rm d}f,
\label{eq:post_dm2}
\end{equation}
in which ${\mathcal L}_{\bar{p}}\propto\exp(-\chi^2_{\bar{p}}/2)$ 
is the likelihood of the model given the AMS-02 antiproton data, 
$\boldsymbol{\theta}_{\rm bkg}$ is the background parameters 
as listed in Table~\ref{table:proton}, $\kappa$ is a constant factor
characterizing the uncertainties of the production cross section of 
antiprotons in $pp$ collisions, $\bar{\phi}_{\rm DM}$ is the flux of 
the DM component calculated with the mean background parameters 
$\bar{\boldsymbol{\theta}}_{\rm bkg}$, $f$ is a constant scale factor 
describing the variation of the fluxes due to the uncertainties of the 
background parameters, $p(\boldsymbol{\theta}_{\rm bkg})$, $p(\kappa)$, 
and $p(f)$ are the prior probabilities of these parameters, respectively.
The prior distribution $p(\boldsymbol{\theta}_{\rm bkg})$ is obtained
through fitting to the proton fluxes (Section 2). To avoid unphysical
results with negative coefficients, we limit the prior regions of $f$
in $[0.2,1.8]$, and $\kappa$ in $[0.5,1.5]$ in the integration. 
We have tested that this approximation gives very similar results as 
the full computation as done in Ref.~\cite{2017PhRvL.118s1101C}.
 
\begin{figure}[!htb]
\centering
\includegraphics[width=0.7\textwidth]{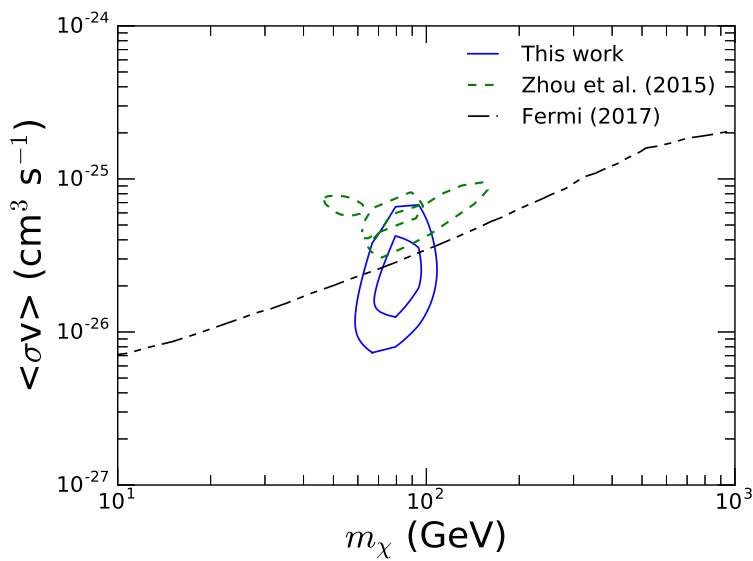}
\caption{Favored $68\%$ and $95\%$ credible regions (solid contours) on 
the mass and annihilation cross section of DM for $b\bar{b}$ channel, 
compared with that inferred from the Galactic center $\gamma$-ray excess
with a re-normalization of the local density (dashed contours; 
\cite{2015PhRvD..91l3010Z}). The $95\%$ exclusion limit of Fermi-LAT 
observations of dwarf galaxies is shown by the dash-dotted line 
\cite{2017ApJ...834..110A}.
\label{fig:contour}}
\end{figure}

We find that a DM component is favored by the AMS-02 data. Assuming
$b\bar{b}$ annihilation channel, the favored mass range of the DM
particles is about $60 \sim 100$ GeV, and the annihilation cross
section is $(0.7 \sim 7)\times10^{-26}$ cm$^3$~s$^{-1}$, as shown in
Figure~\ref{fig:contour}. We estimate the Bayes factor of a DM
component with $b\bar{b}$ annihilation channel is about $8.4$, which 
can be regarded as {\it strong} evidence supporting the DM model. 
These results are consistent with that found previously 
\cite{2017PhRvL.118s1102C,2017PhRvL.118s1101C}, although the Bayes
factor is slightly smaller. The difference of the Bayes factor is
due to the update of the propagation model and parameters with a
more consistent treatment of the fit to the CR data, in particular
the inclusion of the Voyager data. The favored parameter region of DM 
is also consistent with that inferred from the GeV $\gamma$-ray excess 
from the Galactic center (dashed contours\footnote{Different contours 
are due to different assumptions of the diffuse background emission}; 
\cite{2015PhRvD..91l3010Z}). A large part of the favored parameter 
region lies below the constraints 
obtained with Fermi-LAT observations of a class of dwarf galaxies 
(dash-dotted line) \cite{2017ApJ...834..110A}. A global fit to the 
data of antiprotons, and $\gamma$-rays from the Galactic center and 
dwarf galaxies give similar results \cite{2017JCAP...10..053C}, 
although the propagation model and parameters are different from ours. 
It can further be noted that such a cross section is also consistent 
with the value suggested by the relic density for the thermal 
production scenario of DM.

\begin{figure}[!htb]
\centering
\includegraphics[width=0.48\textwidth]{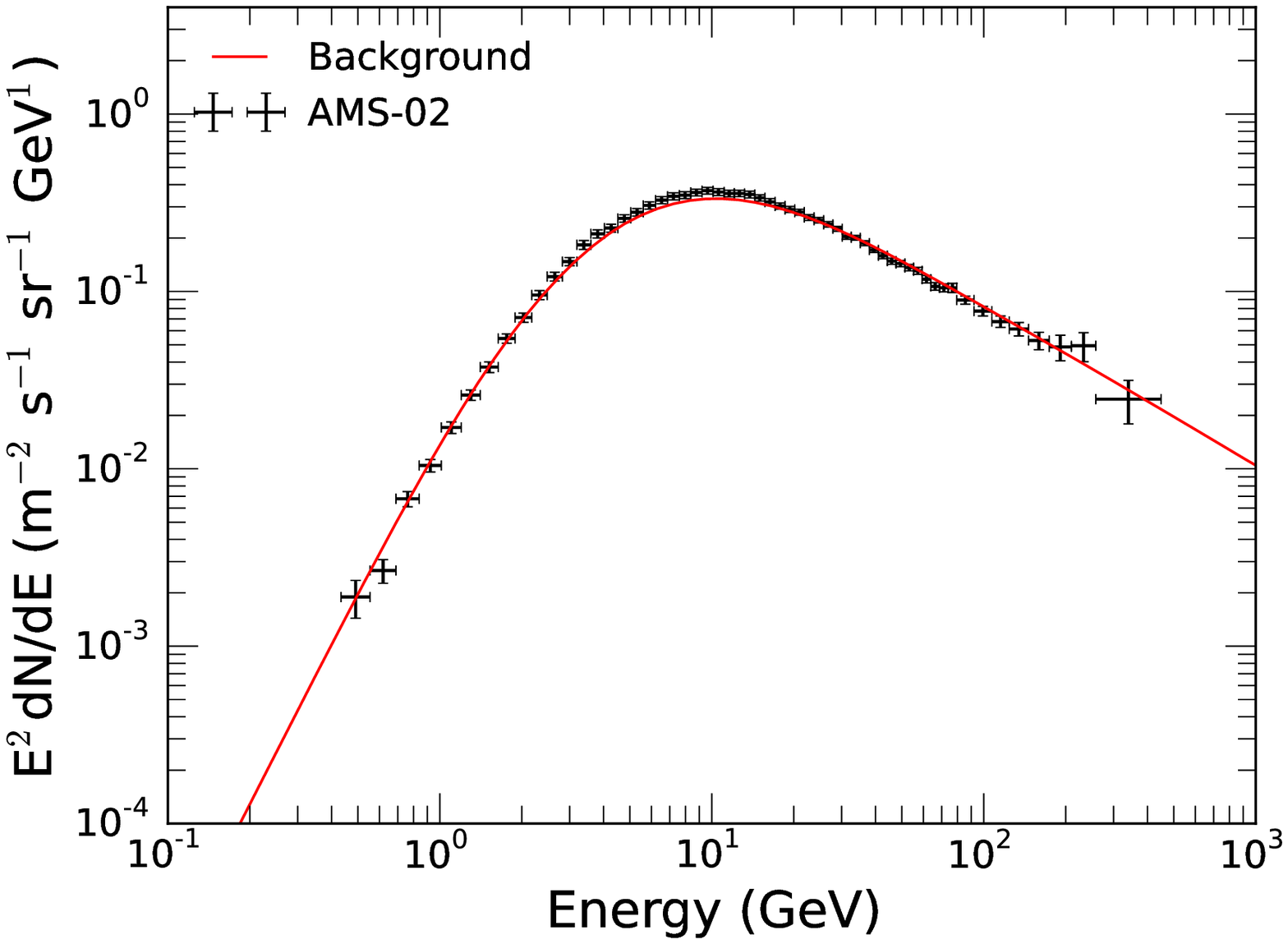}
\includegraphics[width=0.48\textwidth]{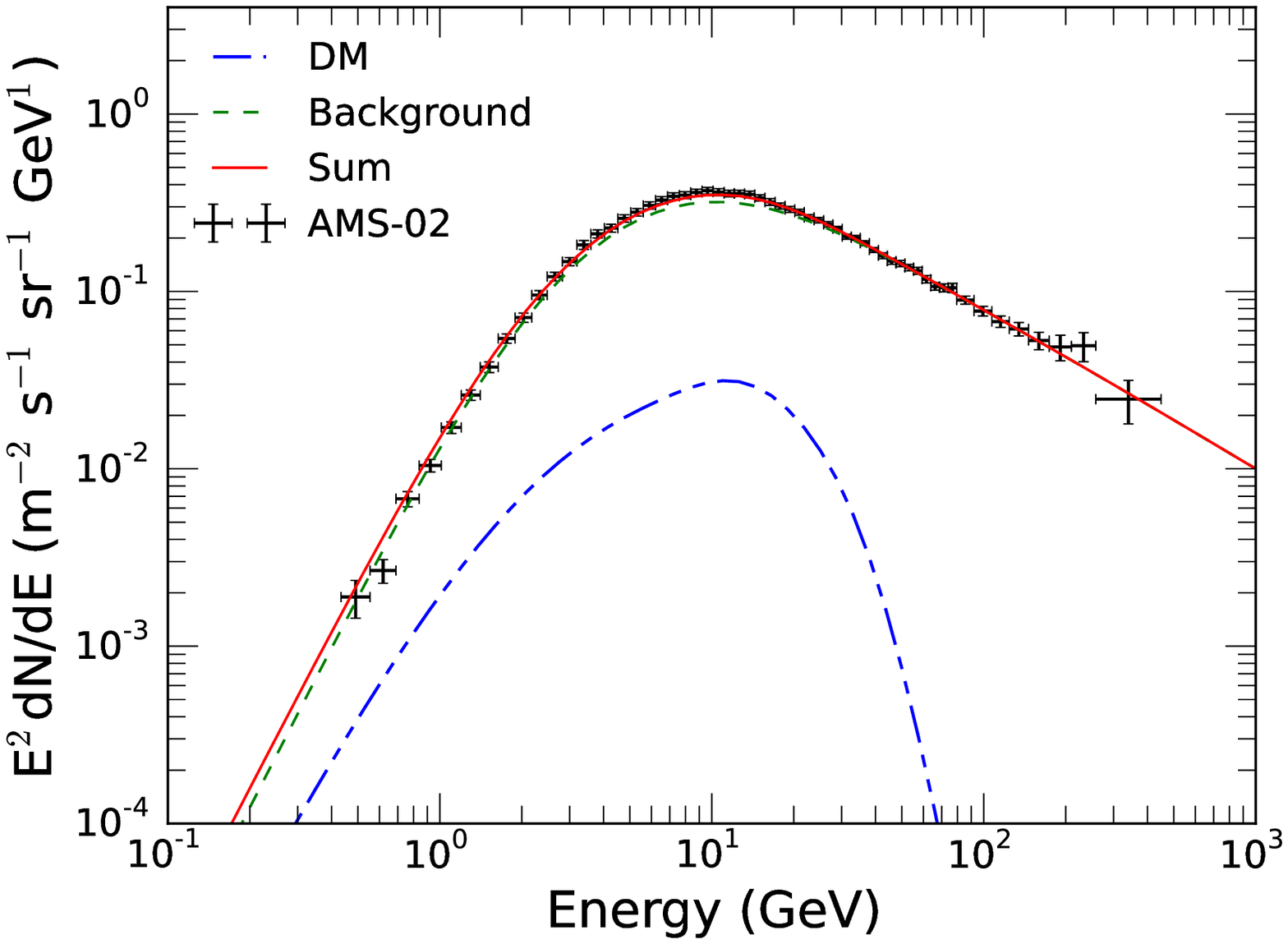}
\caption{Best-fit model predictions of the antiproton fluxes, compared
with the AMS-02 data. The left panel is for the background-only fit, 
and the right one is for the background + DM fit. 
\label{fig:spectrum}}
\end{figure}

Figure~\ref{fig:spectrum} illustrates the antiproton fluxes of the
best-fit models, for the background-only hypothesis (left) and the 
background + DM hypothesis (right), respectively. Compared with the
background prediction, excesses of antiprotons can be seen around a
few GeV. It is interesting to note that at high energies the 
background model is quite consistent with the data, without any 
significant excess \cite{2017PhRvD..95f3021H,2017JHEP...04..112L,
2017PhRvD..95l3007C,2017PhRvD..96l3010L,2017arXiv170102263F}.

\begin{figure}[!htb]
\centering
\includegraphics[width=0.7\textwidth]{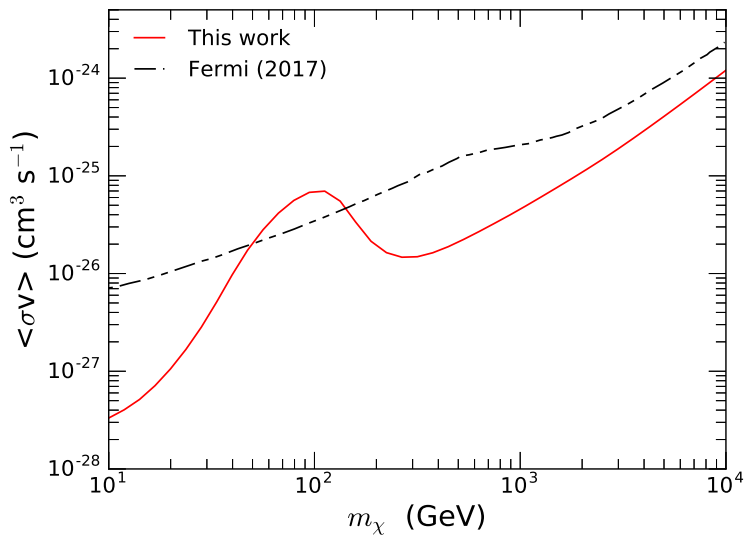}
\caption{$95\%$ credible level upper limits of the annihilation cross
section for $\chi\chi\to b\bar{b}$ obtained from AMS-02 antiproton
data, and that from Fermi-LAT observations of dwarf galaxies
\cite{2017ApJ...834..110A}.
\label{fig:limit}}
\end{figure}

Finally we derive the upper limits on the DM annihilation cross section
from the fit to the antiproton data. The $95\%$ upper limit for given 
$m_{\chi}$ is obtained from the following equation
\begin{equation}
\frac{\int_0^{\sv_{95}}{\mathcal P}(x)\,{\rm d}x}
{\int_0^{\infty}{\mathcal P}(x)\,{\rm d}x}=0.95.
\end{equation}
The results are given in Figure~\ref{fig:limit}. The limits are typically 
stronger than that obtained by $\gamma$-ray observations of dwarf galaxies, 
except for the ``signal region'' with $m_{\chi}\sim50-130$ GeV. These 
limits may scale down by a constant factor, if the local density of DM 
is higher (e.g., \cite{2010JCAP...08..004C,2010A&A...523A..83S}).

\section{Conclusion and discussion}

In this work we revisit the indirect detection of DM with CR antiprotons.
Several important updates are presented. The propagation parameters of
CRs are obtained through fitting to the newest B/C and Carbon flux
data by AMS-02, as well as the low energy observations near the Earth
by ACE and in the local interstellar space by Voyager. Using the 
propagation parameters as priors, we fit the Voyager, AMS-02, and CREAM
data of proton fluxes to obtain the source parameters of CR protons and 
the solar modulation potential. These newly obtained background
parameters are then used to calculate the background antiproton fluxes.
The production cross section of $pp$ collisions based on the recent 
collider data is employed. A Bayesian approach taking into account the
uncertainties of the background parameters and antiproton production 
cross section is adopted.

We find that the AMS-02 antiproton data favor a DM component with mass
of $60 \sim 100$ GeV and annihilation cross section of $(0.7 \sim 7)
\times10^{-26}$ cm$^3$~s$^{-1}$, for an assumed $b\bar{b}$ channel.
The Bayes factor of the DM component is about 8.4. These results are
consistent with previous works based on different propagation
parameters and antiproton production cross sections 
\cite{2017PhRvL.118s1102C,2017PhRvL.118s1101C}.

We discuss possible caveats of the current study. One thing that needs 
to be kept in mind is the uncertainty of the nuclear fragmentation 
cross section, which would affect the calculation of the B/C 
ratio and hence the propagation parameters~\cite{2017arXiv171109616E,
2017PhRvD..96j3005T}. Better measurements of the fragmentation cross
sections for the most relevant species are necessary in future to reduce
such uncertainties. The production cross section of antiprotons also
needs improvements from future experiments. Second, we assume a relatively 
simple propagation paradigm with a uniform diffusion coefficient. 
However, the actual case might be more complicated, as indicated by 
precise measurements of the CR spectra and diffuse $\gamma$-rays
\cite{2016PhRvD..94l3007F,2016ApJ...819...54G,2017JCAP...10..019C}.
The impact of these alternative propagation configurations on the
antiproton calculation deserves further detailed investigation.
Finally, we have assumed that the solar modulation effects of protons 
and antiprotons are similar, with the only difference of the data-taking 
time which corresponds to different solar activities. It is, however, 
possible that the solar modulation is charge-sign-dependent 
\cite{1996ApJ...464..507C,2012AdSpR..49.1587D,2013PhRvL.110h1101M,
2014SoPh..289..391P,2016CoPhC.207..386K}. 
We expect that better understanding of the solar modulation can be
achieved via long-term measurements of both the proton and antiproton
spectra in a full solar cycle. Our understandings about all these 
uncertainties are expected to be improved considerably in the near
future with the continuous operation of CR experiments and the efforts
from colliders, till then the DM contribution to antiprotons can be 
crucially tested.

\acknowledgments
This work is supported by the National Key Research and Development Program 
of China (No. 2016YFA0400200), the National Natural Science Foundation of 
China (Nos. 11475085, 11525313, 11535005, 11690030, 11722328), and the 
100 Talents program of Chinese Academy of Sciences.

\section*{Appendix: Covariance matrix of the propagation parameters 
from fitting to the AMS-02 B/C ratio and Carbon flux}

Here we present the covariance matrix of the propagation parameters,
$\boldsymbol{\theta}=(D_0,\,\delta,\,z_h,\,v_A,\,\eta)$, derived through 
fitting to the B/C ratio and Carbon flux data reported by AMS-02. It is
\begin{equation}
{\rm \Sigma}=\left(
\begin{array}{ccccc}
0.772 & -2.549\times10^{-3} & 0.971 & -0.036 & 0.020 \\
-2.549\times10^{-3} & 7.262\times10^{-5} & -1.355\times10^{-3} & -5.424\times10^{-3} & -5.776\times10^{-4} \\
0.971 & -1.355\times10^{-3} & 1.298 & -0.287 & 0.011 \\
-0.036 & -5.424\times10^{-3} & -0.287 & 1.244 & 0.028 \\
0.020 & -5.776\times10^{-4} & 0.011 & 0.028 & 9.253\times10^{-3}
\end{array}\right).
\nonumber
\end{equation}
This covariance matrix is used in the fit of the proton injection spectrum
as priors. Specifically, we add the following term in the calculation of
the $\chi^2$ of protons
\begin{equation}
\chi^2_{\Sigma}=\left(\boldsymbol{\theta}-\bar{\boldsymbol{\theta}}\right)~
\Sigma^{-1}~\left(\boldsymbol{\theta}-\bar{\boldsymbol{\theta}}\right)^{T},
\end{equation}
where $\bar{\boldsymbol{\theta}}=(6.46,\,0.410,\,6.11,\,29.4,\,-0.48)$ 
is the vector of the mean values of the propagation parameter
\cite{2018arXiv180510649Y}. Note that these values are obtained through 
fitting to the B/C ratio and Carbon flux data, and are slightly different 
from that given in Table~\ref{table:proton}.

\providecommand{\href}[2]{#2}\begingroup\raggedright\endgroup


\begin{thebibliography}{10}

\bibitem{1996PhRvL..76.3057M}
J.~W. {Mitchell}, L.~M. {Barbier}, E.~R. {Christian}, J.~F. {Krizmanic},
  K.~{Krombel}, J.~F. {Ormes} et~al., \emph{{Measurement of 0.25-3.2 GeV
  Antiprotons in the Cosmic Radiation}},
  \href{http://dx.doi.org/10.1103/PhysRevLett.76.3057}{\emph{\prl} {\bf 76}
  (Apr., 1996) 3057--3060}.

\bibitem{1997ApJ...487..415B}
M.~{Boezio}, P.~{Carlson}, T.~{Francke}, N.~{Weber}, M.~{Suffert}, M.~{Hof}
  et~al., \emph{{The Cosmic-Ray Antiproton Flux between 0.62 and 3.19 G eV
  Measured Near Solar Minimum Activity}},
  \href{http://dx.doi.org/10.1086/304593}{\emph{\apj} {\bf 487} (Sept., 1997)
  415}.

\bibitem{1999ICRC....3...77B}
G.~{Basini}, \emph{{The Flux of Cosmic Ray Antiprotons from 3.7 to 24 GeV}},
  {\emph{International Cosmic Ray Conference} {\bf 3} (Aug., 1999) 77}.

\bibitem{2000PhRvL..84.1078O}
S.~{Orito}, T.~{Maeno}, H.~{Matsunaga} and {et al.}, \emph{{Precision
  Measurement of Cosmic-Ray Antiproton Spectrum}}, {\emph{\prl} {\bf 84} (Feb.,
  2000) 1078--1081}, [\href{http://arxiv.org/abs/astro-ph/9906426}{{\tt
  astro-ph/9906426}}].

\bibitem{2001APh....16..121M}
T.~{Maeno}, S.~{Orito}, H.~{Matsunaga}, K.~{Abe}, K.~{Anraku}, Y.~{Asaoka}
  et~al., \emph{{Successive measurements of cosmic-ray antiproton spectrum in a
  positive phase of the solar cycle}},
  \href{http://dx.doi.org/10.1016/S0927-6505(01)00107-4}{\emph{Astroparticle
  Physics} {\bf 16} (Nov., 2001) 121--128},
  [\href{http://arxiv.org/abs/astro-ph/0010381}{{\tt astro-ph/0010381}}].

\bibitem{2001ApJ...561..787B}
M.~{Boezio}, V.~{Bonvicini}, P.~{Schiavon}, A.~{Vacchi}, N.~{Zampa},
  D.~{Bergstr{\"o}m} et~al., \emph{{The Cosmic-Ray Antiproton Flux between 3
  and 49 GeV}}, \href{http://dx.doi.org/10.1086/323366}{\emph{\apj} {\bf 561}
  (Nov., 2001) 787--799}, [\href{http://arxiv.org/abs/astro-ph/0103513}{{\tt
  astro-ph/0103513}}].

\bibitem{2001PhRvL..87A1101B}
A.~S. {Beach}, J.~J. {Beatty}, A.~{Bhattacharyya}, C.~{Bower}, S.~{Coutu},
  M.~A. {Duvernois} et~al., \emph{{Measurement of the Cosmic-Ray
  Antiproton-to-Proton Abundance Ratio between 4 and 50 GeV}},
  \href{http://dx.doi.org/10.1103/PhysRevLett.87.271101}{\emph{\prl} {\bf 87}
  (Dec., 2001) 261101}, [\href{http://arxiv.org/abs/astro-ph/0111094}{{\tt
  astro-ph/0111094}}].

\bibitem{2002PhR...366..331A}
M.~{Aguilar}, J.~{Alcaraz}, J.~{Allaby}, B.~{Alpat}, G.~{Ambrosi},
  H.~{Anderhub} et~al., \emph{{The Alpha Magnetic Spectrometer (AMS) on the
  International Space Station: Part I - results from the test flight on the
  space shuttle}},
  \href{http://dx.doi.org/10.1016/S0370-1573(02)00013-3}{\emph{\physrep} {\bf
  366} (Aug., 2002) 331--405}.

\bibitem{2002PhRvL..88e1101A}
Y.~{Asaoka}, Y.~{Shikaze}, K.~{Abe}, K.~{Anraku}, M.~{Fujikawa}, H.~{Fuke}
  et~al., \emph{{Measurements of Cosmic-Ray Low-Energy Antiproton and Proton
  Spectra in a Transient Period of Solar Field Reversal}},
  \href{http://dx.doi.org/10.1103/PhysRevLett.88.051101}{\emph{\prl} {\bf 88}
  (Feb., 2002) 051101}, [\href{http://arxiv.org/abs/astro-ph/0109007}{{\tt
  astro-ph/0109007}}].

\bibitem{2005ICRC....3...13H}
S.~{Haino}, K.~{Abe}, H.~{Fuke}, T.~{Maeno}, Y.~{Makida}, H.~{Matsumoto}
  et~al., \emph{{Measurement of cosmic-ray antiproton spectrum with
  BESS-2002}},  in \emph{International Cosmic Ray Conference}, vol.~3, p.~13,
  2005.

\bibitem{2008PhLB..670..103A}
K.~{Abe}, H.~{Fuke}, S.~{Haino}, T.~{Hams}, A.~{Itazaki}, K.~C. {Kim} et~al.,
  \emph{{Measurement of the cosmic-ray low-energy antiproton spectrum with the
  first BESS-Polar Antarctic flight}},
  \href{http://dx.doi.org/10.1016/j.physletb.2008.10.053}{\emph{Physics Letters
  B} {\bf 670} (Dec., 2008) 103--108},
  [\href{http://arxiv.org/abs/0805.1754}{{\tt 0805.1754}}].

\bibitem{2009PhRvL.102e1101A}
O.~{Adriani}, G.~C. {Barbarino}, G.~A. {Bazilevskaya}, R.~{Bellotti},
  M.~{Boezio}, E.~A. {Bogomolov} et~al., \emph{{New Measurement of the
  Antiproton-to-Proton Flux Ratio up to 100 GeV in the Cosmic Radiation}},
  \href{http://dx.doi.org/10.1103/PhysRevLett.102.051101}{\emph{\prl} {\bf 102}
  (Feb., 2009) 051101}, [\href{http://arxiv.org/abs/0810.4994}{{\tt
  0810.4994}}].

\bibitem{2010PhRvL.105l1101A}
O.~{Adriani}, G.~C. {Barbarino}, G.~A. {Bazilevskaya}, R.~{Bellotti},
  M.~{Boezio}, E.~A. {Bogomolov} et~al., \emph{{PAMELA Results on the
  Cosmic-Ray Antiproton Flux from 60 MeV to 180 GeV in Kinetic Energy}},
  \href{http://dx.doi.org/10.1103/PhysRevLett.105.121101}{\emph{\prl} {\bf 105}
  (Sept., 2010) 121101}, [\href{http://arxiv.org/abs/1007.0821}{{\tt
  1007.0821}}].

\bibitem{2016PhRvL.117i1103A}
M.~{Aguilar}, L.~{Ali Cavasonza}, B.~{Alpat}, G.~{Ambrosi}, L.~{Arruda},
  N.~{Attig} et~al., \emph{{Antiproton Flux, Antiproton-to-Proton Flux Ratio,
  and Properties of Elementary Particle Fluxes in Primary Cosmic Rays Measured
  with the Alpha Magnetic Spectrometer on the International Space Station}},
  \href{http://dx.doi.org/10.1103/PhysRevLett.117.091103}{\emph{\prl} {\bf 117}
  (Aug., 2016) 091103}.

\bibitem{2018arXiv180110310A}
O.~{Adriani}, G.~C. {Barbarino}, G.~A. {Bazilevskaya}, R.~{Bellotti},
  M.~{Boezio}, E.~A. {Bogomolov} et~al., \emph{{Ten Years of PAMELA in Space}},
  {\emph{ArXiv e-prints} (Jan., 2018) },
  [\href{http://arxiv.org/abs/1801.10310}{{\tt 1801.10310}}].

\bibitem{2016PhRvD..94l3019K}
M.~{Korsmeier} and A.~{Cuoco}, \emph{{Galactic cosmic-ray propagation in the
  light of AMS-02: Analysis of protons, helium, and antiprotons}},
  \href{http://dx.doi.org/10.1103/PhysRevD.94.123019}{\emph{\prd} {\bf 94}
  (Dec., 2016) 123019}, [\href{http://arxiv.org/abs/1607.06093}{{\tt
  1607.06093}}].

\bibitem{2016ApJ...824...16J}
G.~{J{\'o}hannesson}, R.~{Ruiz de Austri}, A.~C. {Vincent}, I.~V. {Moskalenko},
  E.~{Orlando}, T.~A. {Porter} et~al., \emph{{Bayesian Analysis of Cosmic Ray
  Propagation: Evidence against Homogeneous Diffusion}},
  \href{http://dx.doi.org/10.3847/0004-637X/824/1/16}{\emph{\apj} {\bf 824}
  (June, 2016) 16}, [\href{http://arxiv.org/abs/1602.02243}{{\tt 1602.02243}}].

\bibitem{2015PhRvD..92e5027J}
H.-B. {Jin}, Y.-L. {Wu} and Y.-F. {Zhou}, \emph{{Upper limits on dark matter
  annihilation cross sections from the first AMS-02 antiproton data}},
  \href{http://dx.doi.org/10.1103/PhysRevD.92.055027}{\emph{\prd} {\bf 92}
  (Sept., 2015) 055027}, [\href{http://arxiv.org/abs/1504.04604}{{\tt
  1504.04604}}].

\bibitem{2017PhRvD..96l3010L}
S.-J. {Lin}, X.-J. {Bi}, J.~{Feng}, P.-F. {Yin} and Z.-H. {Yu},
  \emph{{Systematic study on the cosmic ray antiproton flux}},
  \href{http://dx.doi.org/10.1103/PhysRevD.96.123010}{\emph{\prd} {\bf 96}
  (Dec., 2017) 123010}, [\href{http://arxiv.org/abs/1612.04001}{{\tt
  1612.04001}}].

\bibitem{2016PTEP.2016b1E01K}
K.~{Kohri}, K.~{Ioka}, Y.~{Fujita} and R.~{Yamazaki}, \emph{{Can we explain
  AMS-02 antiproton and positron excesses simultaneously by nearby supernovae
  without pulsars or dark matter?}},
  \href{http://dx.doi.org/10.1093/ptep/ptv193}{\emph{Progress of Theoretical
  and Experimental Physics} {\bf 2016} (Feb., 2016) 021E01},
  [\href{http://arxiv.org/abs/1505.01236}{{\tt 1505.01236}}].

\bibitem{2017PhRvD..96b3006L}
W.~{Liu}, X.-J. {Bi}, S.-J. {Lin}, B.-B. {Wang} and P.-F. {Yin},
  \emph{{Excesses of cosmic ray spectra from a single nearby source}},
  \href{http://dx.doi.org/10.1103/PhysRevD.96.023006}{\emph{\prd} {\bf 96}
  (July, 2017) 023006}, [\href{http://arxiv.org/abs/1611.09118}{{\tt
  1611.09118}}].

\bibitem{2017PhRvD..95l3007C}
I.~{Cholis}, D.~{Hooper} and T.~{Linden}, \emph{{Possible evidence for the
  stochastic acceleration of secondary antiprotons by supernova remnants}},
  \href{http://dx.doi.org/10.1103/PhysRevD.95.123007}{\emph{\prd} {\bf 95}
  (June, 2017) 123007}, [\href{http://arxiv.org/abs/1701.04406}{{\tt
  1701.04406}}].

\bibitem{2017PhRvD..95f3021H}
X.-J. {Huang}, C.-C. {Wei}, Y.-L. {Wu}, W.-H. {Zhang} and Y.-F. {Zhou},
  \emph{{Antiprotons from dark matter annihilation through light mediators and
  a possible excess in AMS-02 $\bar{p}$/p data}},
  \href{http://dx.doi.org/10.1103/PhysRevD.95.063021}{\emph{\prd} {\bf 95}
  (Mar., 2017) 063021}, [\href{http://arxiv.org/abs/1611.01983}{{\tt
  1611.01983}}].

\bibitem{2017JHEP...04..112L}
T.~{Li}, \emph{{Simplified dark matter models in the light of AMS-02 antiproton
  data}}, \href{http://dx.doi.org/10.1007/JHEP04(2017)112}{\emph{Journal of
  High Energy Physics} {\bf 04} (Apr., 2017) 112},
  [\href{http://arxiv.org/abs/1612.09501}{{\tt 1612.09501}}].

\bibitem{2017arXiv170102263F}
J.~{Feng} and H.-H. {Zhang}, \emph{{Dark Matter Search in Space: Combined
  Analysis of Cosmic Ray Antiproton-to-Proton Flux Ratio and Positron Flux
  Measured by AMS-02}}, {\emph{ArXiv e-prints} (Jan., 2017) },
  [\href{http://arxiv.org/abs/1701.02263}{{\tt 1701.02263}}].

\bibitem{2001ApJ...563..172D}
F.~{Donato}, D.~{Maurin}, P.~{Salati}, A.~{Barrau}, G.~{Boudoul} and
  R.~{Taillet}, \emph{{Antiprotons from Spallations of Cosmic Rays on
  Interstellar Matter}}, {\emph{\apj} {\bf 563} (Dec., 2001) 172--184}.

\bibitem{2004PhRvD..69f3501D}
F.~{Donato}, N.~{Fornengo}, D.~{Maurin}, P.~{Salati} and R.~{Taillet},
  \emph{{Antiprotons in cosmic rays from neutralino annihilation}},
  \href{http://dx.doi.org/10.1103/PhysRevD.69.063501}{\emph{\prd} {\bf 69}
  (Mar., 2004) 063501}, [\href{http://arxiv.org/abs/astro-ph/0306207}{{\tt
  astro-ph/0306207}}].

\bibitem{2015JCAP...03..021H}
D.~{Hooper}, T.~{Linden} and P.~{Mertsch}, \emph{{What does the PAMELA
  antiproton spectrum tell us about dark matter?}},
  \href{http://dx.doi.org/10.1088/1475-7516/2015/03/021}{\emph{\jcap} {\bf 3}
  (Mar., 2015) 21}, [\href{http://arxiv.org/abs/1410.1527}{{\tt 1410.1527}}].

\bibitem{2001ApJ...555..585M}
D.~{Maurin}, F.~{Donato}, R.~{Taillet} and P.~{Salati}, \emph{{Cosmic Rays
  below Z=30 in a Diffusion Model: New Constraints on Propagation Parameters}},
  {\emph{\apj} {\bf 555} (July, 2001) 585--596}.

\bibitem{2011ApJ...729..106T}
R.~{Trotta}, G.~{J{\'o}hannesson}, I.~V. {Moskalenko}, T.~A. {Porter}, R.~{Ruiz
  de Austri} and A.~W. {Strong}, \emph{{Constraints on Cosmic-ray Propagation
  Models from A Global Bayesian Analysis}},
  \href{http://dx.doi.org/10.1088/0004-637X/729/2/106}{\emph{\apj} {\bf 729}
  (Mar., 2011) 106}, [\href{http://arxiv.org/abs/1011.0037}{{\tt 1011.0037}}].

\bibitem{2015JCAP...09..049J}
H.-B. {Jin}, Y.-L. {Wu} and Y.-F. {Zhou}, \emph{{Cosmic ray propagation and
  dark matter in light of the latest AMS-02 data}},
  \href{http://dx.doi.org/10.1088/1475-7516/2015/09/049}{\emph{\jcap} {\bf 9}
  (Sept., 2015) 49}, [\href{http://arxiv.org/abs/1410.0171}{{\tt 1410.0171}}].

\bibitem{2016PhRvD..94l3007F}
J.~{Feng}, N.~{Tomassetti} and A.~{Oliva}, \emph{{Bayesian analysis of
  spatial-dependent cosmic-ray propagation: Astrophysical background of
  antiprotons and positrons}},
  \href{http://dx.doi.org/10.1103/PhysRevD.94.123007}{\emph{\prd} {\bf 94}
  (Dec., 2016) 123007}, [\href{http://arxiv.org/abs/1610.06182}{{\tt
  1610.06182}}].

\bibitem{2016PhRvL.117w1102A}
M.~{Aguilar}, D.~{Aisa}, A.~{Alvino}, G.~{Ambrosi}, K.~{Andeen}, L.~{Arruda}
  et~al., \emph{{Precise Measurement of the Boron to Carbon Flux Ratio in
  Cosmic Rays from 1.9 GV to 2.6 TV with the Alpha Magnetic Spectrometer on the
  International Space Station}},
  \href{http://dx.doi.org/10.1103/PhysRevLett.117.231102}{\emph{\prl} {\bf 117}
  (Nov., 2016) 231102}.

\bibitem{2017PhRvD..95h3007Y}
Q.~{Yuan}, S.-J. {Lin}, K.~{Fang} and X.-J. {Bi}, \emph{{Propagation of cosmic
  rays in the AMS-02 era}},
  \href{http://dx.doi.org/10.1103/PhysRevD.95.083007}{\emph{\prd} {\bf 95}
  (Apr., 2017) 083007}, [\href{http://arxiv.org/abs/1701.06149}{{\tt
  1701.06149}}].

\bibitem{2018PhRvD..97b3015N}
J.-S. {Niu} and T.~{Li}, \emph{{Galactic cosmic-ray model in the light of
  AMS-02 nuclei data}},
  \href{http://dx.doi.org/10.1103/PhysRevD.97.023015}{\emph{\prd} {\bf 97}
  (Jan., 2018) 023015}, [\href{http://arxiv.org/abs/1705.11089}{{\tt
  1705.11089}}].

\bibitem{2018JCAP...01..055R}
A.~{Reinert} and M.~W. {Winkler}, \emph{{A precision search for WIMPs with
  charged cosmic rays}},
  \href{http://dx.doi.org/10.1088/1475-7516/2018/01/055}{\emph{\jcap} {\bf 1}
  (Jan., 2018) 055}, [\href{http://arxiv.org/abs/1712.00002}{{\tt
  1712.00002}}].

\bibitem{2017PhRvL.118s1101C}
M.-Y. {Cui}, Q.~{Yuan}, Y.-L. {Sming Tsai} and Y.-Z. {Fan}, \emph{{A possible
  dark matter annihilation signal in the AMS-02 antiproton data}},
  \href{http://dx.doi.org/10.1103/PhysRevLett.118.191101}{\emph{\prl} {\bf 118}
  (May, 2017) 191101}, [\href{http://arxiv.org/abs/1610.03840}{{\tt
  1610.03840}}].

\bibitem{2017PhRvL.118s1102C}
A.~{Cuoco}, M.~{Kramer} and M.~{Korsmeier}, \emph{{Novel dark matter
  constraints from antiprotons in the light of AMS-02}},
  \href{http://dx.doi.org/10.1103/PhysRevLett.118.191102}{\emph{\prl} {\bf 118}
  (May, 2017) 191102}, [\href{http://arxiv.org/abs/1610.03071}{{\tt
  1610.03071}}].

\bibitem{2011PhLB..697..412H}
D.~{Hooper} and L.~{Goodenough}, \emph{{Dark matter annihilation in the
  Galactic Center as seen by the Fermi Gamma Ray Space Telescope}},
  \href{http://dx.doi.org/10.1016/j.physletb.2011.02.029}{\emph{Physics Letters
  B} {\bf 697} (Mar., 2011) 412--428},
  [\href{http://arxiv.org/abs/1010.2752}{{\tt 1010.2752}}].

\bibitem{2012PhRvD..86h3511A}
K.~N. {Abazajian} and M.~{Kaplinghat}, \emph{{Detection of a gamma-ray source
  in the Galactic Center consistent with extended emission from dark matter
  annihilation and concentrated astrophysical emission}},
  \href{http://dx.doi.org/10.1103/PhysRevD.86.083511}{\emph{\prd} {\bf 86}
  (Oct., 2012) 083511}, [\href{http://arxiv.org/abs/1207.6047}{{\tt
  1207.6047}}].

\bibitem{2015PhRvD..91l3010Z}
B.~{Zhou}, Y.-F. {Liang}, X.~{Huang}, X.~{Li}, Y.-Z. {Fan}, L.~{Feng} et~al.,
  \emph{{GeV excess in the Milky Way: The role of diffuse galactic gamma-ray
  emission templates}},
  \href{http://dx.doi.org/10.1103/PhysRevD.91.123010}{\emph{\prd} {\bf 91}
  (June, 2015) 123010}, [\href{http://arxiv.org/abs/1406.6948}{{\tt
  1406.6948}}].

\bibitem{2016JCAP...04..030H}
X.~{Huang}, T.~{En{\ss}lin} and M.~{Selig}, \emph{{Galactic dark matter search
  via phenomenological astrophysics modeling}},
  \href{http://dx.doi.org/10.1088/1475-7516/2016/04/030}{\emph{\jcap} {\bf 4}
  (Apr., 2016) 030}, [\href{http://arxiv.org/abs/1511.02621}{{\tt
  1511.02621}}].

\bibitem{2015PhRvL.115h1101G}
A.~{Geringer-Sameth}, M.~G. {Walker}, S.~M. {Koushiappas}, S.~E. {Koposov},
  V.~{Belokurov}, G.~{Torrealba} et~al., \emph{{Indication of Gamma-Ray
  Emission from the Newly Discovered Dwarf Galaxy Reticulum II}},
  \href{http://dx.doi.org/10.1103/PhysRevLett.115.081101}{\emph{\prl} {\bf 115}
  (Aug., 2015) 081101}, [\href{http://arxiv.org/abs/1503.02320}{{\tt
  1503.02320}}].

\bibitem{2016PhRvD..93d3518L}
S.~{Li}, Y.-F. {Liang}, K.-K. {Duan}, Z.-Q. {Shen}, X.~{Huang}, X.~{Li} et~al.,
  \emph{{Search for gamma-ray emission from eight dwarf spheroidal galaxy
  candidates discovered in year two of Dark Energy Survey with Fermi-LAT
  data}}, \href{http://dx.doi.org/10.1103/PhysRevD.93.043518}{\emph{\prd} {\bf
  93} (Feb., 2016) 043518}, [\href{http://arxiv.org/abs/1511.09252}{{\tt
  1511.09252}}].

\bibitem{2016PhRvD..93j3525L}
Y.-F. {Liang}, Z.-Q. {Shen}, X.~{Li}, Y.-Z. {Fan}, X.~{Huang}, S.-J. {Lei}
  et~al., \emph{{Search for a gamma-ray line feature from a group of nearby
  galaxy clusters with Fermi LAT Pass 8 data}},
  \href{http://dx.doi.org/10.1103/PhysRevD.93.103525}{\emph{\prd} {\bf 93}
  (May, 2016) 103525}, [\href{http://arxiv.org/abs/1602.06527}{{\tt
  1602.06527}}].

\bibitem{2017JCAP...10..053C}
A.~{Cuoco}, J.~{Heisig}, M.~{Korsmeier} and M.~{Kr{\"a}mer}, \emph{{Probing
  dark matter annihilation in the Galaxy with antiprotons and gamma rays}},
  \href{http://dx.doi.org/10.1088/1475-7516/2017/10/053}{\emph{\jcap} {\bf 10}
  (Oct., 2017) 053}, [\href{http://arxiv.org/abs/1704.08258}{{\tt
  1704.08258}}].

\bibitem{2017PhRvL.119y1101A}
M.~{Aguilar}, L.~{Ali Cavasonza}, B.~{Alpat}, G.~{Ambrosi}, L.~{Arruda},
  N.~{Attig} et~al., \emph{{Observation of the Identical Rigidity Dependence of
  He, C, and O Cosmic Rays at High Rigidities by the Alpha Magnetic
  Spectrometer on the International Space Station}},
  \href{http://dx.doi.org/10.1103/PhysRevLett.119.251101}{\emph{\prl} {\bf 119}
  (Dec., 2017) 251101}.

\bibitem{2018PhRvL.120b1101A}
M.~{Aguilar}, L.~{Ali Cavasonza}, B.~{Alpat}, G.~{Ambrosi}, L.~{Arruda},
  N.~{Attig} et~al., \emph{{Observation of New Properties of Secondary Cosmic
  Rays Lithium, Beryllium, and Boron by the Alpha Magnetic Spectrometer on the
  International Space Station}},
  \href{http://dx.doi.org/10.1103/PhysRevLett.120.021101}{\emph{\prl} {\bf 120}
  (Jan., 2018) 021101}.

\bibitem{2013Sci...341..150S}
E.~C. {Stone}, A.~C. {Cummings}, F.~B. {McDonald}, B.~C. {Heikkila}, N.~{Lal}
  and W.~R. {Webber}, \emph{{Voyager 1 Observes Low-Energy Galactic Cosmic Rays
  in a Region Depleted of Heliospheric Ions}},
  \href{http://dx.doi.org/10.1126/science.1236408}{\emph{Science} {\bf 341}
  (July, 2013) 150--153}.

\bibitem{2016ApJ...831...18C}
A.~C. {Cummings}, E.~C. {Stone}, B.~C. {Heikkila}, N.~{Lal}, W.~R. {Webber},
  G.~{J{\'o}hannesson} et~al., \emph{{Galactic Cosmic Rays in the Local
  Interstellar Medium: Voyager 1 Observations and Model Results}},
  \href{http://dx.doi.org/10.3847/0004-637X/831/1/18}{\emph{\apj} {\bf 831}
  (Nov., 2016) 18}.

\bibitem{2018arXiv180510649Y}
Q.~{Yuan}, \emph{{Implications on cosmic ray injection and propagation
  parameters from Voyager/ACE/AMS-02 nucleus data}}, {\emph{ArXiv e-prints}
  (May, 2018) }, [\href{http://arxiv.org/abs/1805.10649}{{\tt 1805.10649}}].

\bibitem{2017JCAP...02..048W}
M.~W. {Winkler}, \emph{{Cosmic ray antiprotons at high energies}},
  \href{http://dx.doi.org/10.1088/1475-7516/2017/02/048}{\emph{\jcap} {\bf 2}
  (Feb., 2017) 048}, [\href{http://arxiv.org/abs/1701.04866}{{\tt
  1701.04866}}].

\bibitem{2016ApJ...819...54G}
Y.-Q. {Guo}, Z.~{Tian} and C.~{Jin}, \emph{{Spatial-dependent Propagation of
  Cosmic Rays Results in the Spectrum of Proton, Ratios of P/P, and B/C, and
  Anisotropy of Nuclei}},
  \href{http://dx.doi.org/10.3847/0004-637X/819/1/54}{\emph{\apj} {\bf 819}
  (Mar., 2016) 54}.

\bibitem{2017JCAP...10..019C}
S.~S. {Cerri}, D.~{Gaggero}, A.~{Vittino}, C.~{Evoli} and D.~{Grasso}, \emph{{A
  signature of anisotropic cosmic-ray transport in the gamma-ray sky}},
  \href{http://dx.doi.org/10.1088/1475-7516/2017/10/019}{\emph{\jcap} {\bf 10}
  (Oct., 2017) 019}, [\href{http://arxiv.org/abs/1707.07694}{{\tt
  1707.07694}}].

\bibitem{2010ApJ...714L..89A}
H.~S. {Ahn}, P.~{Allison}, M.~G. {Bagliesi}, J.~J. {Beatty}, G.~{Bigongiari},
  J.~T. {Childers} et~al., \emph{{Discrepant Hardening Observed in Cosmic-ray
  Elemental Spectra}},
  \href{http://dx.doi.org/10.1088/2041-8205/714/1/L89}{\emph{\apjl} {\bf 714}
  (May, 2010) L89--L93}, [\href{http://arxiv.org/abs/1004.1123}{{\tt
  1004.1123}}].

\bibitem{2011Sci...332...69A}
O.~{Adriani}, G.~C. {Barbarino}, G.~A. {Bazilevskaya}, R.~{Bellotti},
  M.~{Boezio}, E.~A. {Bogomolov} et~al., \emph{{PAMELA Measurements of
  Cosmic-Ray Proton and Helium Spectra}},
  \href{http://dx.doi.org/10.1126/science.1199172}{\emph{Science} {\bf 332}
  (Apr., 2011) 69--}, [\href{http://arxiv.org/abs/1103.4055}{{\tt 1103.4055}}].

\bibitem{2016PhRvD..93l3007Y}
R.~{Yang}, F.~{Aharonian} and C.~{Evoli}, \emph{{Radial distribution of the
  diffuse {$\gamma$} -ray emissivity in the Galactic disk}},
  \href{http://dx.doi.org/10.1103/PhysRevD.93.123007}{\emph{\prd} {\bf 93}
  (June, 2016) 123007}, [\href{http://arxiv.org/abs/1602.04710}{{\tt
  1602.04710}}].

\bibitem{2016ApJS..223...26A}
F.~{Acero}, M.~{Ackermann}, M.~{Ajello}, A.~{Albert}, L.~{Baldini}, J.~{Ballet}
  et~al., \emph{{Development of the Model of Galactic Interstellar Emission for
  Standard Point-source Analysis of Fermi Large Area Telescope Data}},
  \href{http://dx.doi.org/10.3847/0067-0049/223/2/26}{\emph{\apjs} {\bf 223}
  (Apr., 2016) 26}, [\href{http://arxiv.org/abs/1602.07246}{{\tt 1602.07246}}].

\bibitem{1994ApJ...431..705S}
E.~S. {Seo} and V.~S. {Ptuskin}, \emph{{Stochastic reacceleration of cosmic
  rays in the interstellar medium}},
  \href{http://dx.doi.org/10.1086/174520}{\emph{\apj} {\bf 431} (Aug., 1994)
  705--714}.

\bibitem{1968ApJ...154.1011G}
L.~J. {Gleeson} and W.~I. {Axford}, \emph{{Solar Modulation of Galactic Cosmic
  Rays}}, \href{http://dx.doi.org/10.1086/149822}{\emph{\apj} {\bf 154} (Dec.,
  1968) 1011}.

\bibitem{1998ApJ...509..212S}
A.~W. {Strong} and I.~V. {Moskalenko}, \emph{{Propagation of Cosmic-Ray
  Nucleons in the Galaxy}}, \href{http://dx.doi.org/10.1086/306470}{\emph{\apj}
  {\bf 509} (Dec., 1998) 212--228},
  [\href{http://arxiv.org/abs/astro-ph/9807150}{{\tt astro-ph/9807150}}].

\bibitem{1998ApJ...493..694M}
I.~V. {Moskalenko} and A.~W. {Strong}, \emph{{Production and Propagation of
  Cosmic-Ray Positrons and Electrons}},
  \href{http://dx.doi.org/10.1086/305152}{\emph{\apj} {\bf 493} (Jan., 1998)
  694}, [\href{http://arxiv.org/abs/astro-ph/9710124}{{\tt astro-ph/9710124}}].

\bibitem{2010APh....34..274D}
G.~{di Bernardo}, C.~{Evoli}, D.~{Gaggero}, D.~{Grasso} and L.~{Maccione},
  \emph{{Unified interpretation of cosmic ray nuclei and antiproton recent
  measurements}},
  \href{http://dx.doi.org/10.1016/j.astropartphys.2010.08.006}{\emph{Astropart%
icle Physics} {\bf 34} (Dec., 2010) 274--283},
  [\href{http://arxiv.org/abs/0909.4548}{{\tt 0909.4548}}].

\bibitem{2006ApJ...642..902P}
V.~S. {Ptuskin}, I.~V. {Moskalenko}, F.~C. {Jones}, A.~W. {Strong} and V.~N.
  {Zirakashvili}, \emph{{Dissipation of Magnetohydrodynamic Waves on Energetic
  Particles: Impact on Interstellar Turbulence and Cosmic-Ray Transport}},
  \href{http://dx.doi.org/10.1086/501117}{\emph{\apj} {\bf 642} (May, 2006)
  902--916}, [\href{http://arxiv.org/abs/astro-ph/0510335}{{\tt
  astro-ph/0510335}}].

\bibitem{2012PhRvD..85d3507L}
J.~{Liu}, Q.~{Yuan}, X.-J. {Bi}, H.~{Li} and X.~{Zhang}, \emph{{Cosmic ray
  Monte Carlo: A global fitting method in studying the properties of the new
  sources of cosmic e$^{\pm}$ excesses}},
  \href{http://dx.doi.org/10.1103/PhysRevD.85.043507}{\emph{\prd} {\bf 85}
  (Feb., 2012) 043507}, [\href{http://arxiv.org/abs/1106.3882}{{\tt
  1106.3882}}].

\bibitem{2015PhRvL.114q1103A}
M.~{Aguilar}, D.~{Aisa}, B.~{Alpat}, A.~{Alvino}, G.~{Ambrosi}, K.~{Andeen}
  et~al., \emph{{Precision Measurement of the Proton Flux in Primary Cosmic
  Rays from Rigidity 1 GV to 1.8 TV with the Alpha Magnetic Spectrometer on the
  International Space Station}},
  \href{http://dx.doi.org/10.1103/PhysRevLett.114.171103}{\emph{\prl} {\bf 114}
  (May, 2015) 171103}.

\bibitem{2017ApJ...839....5Y}
Y.~S. {Yoon}, T.~{Anderson}, A.~{Barrau}, N.~B. {Conklin}, S.~{Coutu},
  L.~{Derome} et~al., \emph{{Proton and Helium Spectra from the CREAM-III
  Flight}}, \href{http://dx.doi.org/10.3847/1538-4357/aa68e4}{\emph{\apj} {\bf
  839} (Apr., 2017) 5}, [\href{http://arxiv.org/abs/1704.02512}{{\tt
  1704.02512}}].

\bibitem{1983JPhG....9.1289T}
L.~C. {Tan} and L.~K. {Ng}, \emph{{Parametrisation of hadron inclusive cross
  sections in p-p collisions extended to very low energies}},
  \href{http://dx.doi.org/10.1088/0305-4616/9/10/015}{\emph{Journal of Physics
  G Nuclear Physics} {\bf 9} (Oct., 1983) 1289--1308}.

\bibitem{2003PhRvD..68i4017D}
R.~P. {Duperray}, C.-Y. {Huang}, K.~V. {Protasov} and M.~{Bu{\'e}nerd},
  \emph{{Parametrization of the antiproton inclusive production cross section
  on nuclei}}, \href{http://dx.doi.org/10.1103/PhysRevD.68.094017}{\emph{\prd}
  {\bf 68} (Nov., 2003) 094017},
  [\href{http://arxiv.org/abs/astro-ph/0305274}{{\tt astro-ph/0305274}}].

\bibitem{2014PhRvD..90h5017D}
M.~{di Mauro}, F.~{Donato}, A.~{Goudelis} and P.~D. {Serpico}, \emph{{New
  evaluation of the antiproton production cross section for cosmic ray
  studies}}, \href{http://dx.doi.org/10.1103/PhysRevD.90.085017}{\emph{\prd}
  {\bf 90} (Oct., 2014) 085017}, [\href{http://arxiv.org/abs/1408.0288}{{\tt
  1408.0288}}].

\bibitem{2014JCAP...09..051K}
R.~{Kappl} and M.~W. {Winkler}, \emph{{The cosmic ray antiproton background for
  AMS-02}}, \href{http://dx.doi.org/10.1088/1475-7516/2014/09/051}{\emph{\jcap}
  {\bf 9} (Sept., 2014) 051}, [\href{http://arxiv.org/abs/1408.0299}{{\tt
  1408.0299}}].

\bibitem{2015ApJ...803...54K}
M.~{Kachelriess}, I.~V. {Moskalenko} and S.~S. {Ostapchenko}, \emph{{New
  Calculation of Antiproton Production by Cosmic Ray Protons and Nuclei}},
  \href{http://dx.doi.org/10.1088/0004-637X/803/2/54}{\emph{\apj} {\bf 803}
  (Apr., 2015) 54}, [\href{http://arxiv.org/abs/1502.04158}{{\tt 1502.04158}}].

\bibitem{2018arXiv180203030K}
M.~{Korsmeier}, F.~{Donato} and M.~{Di Mauro}, \emph{{Production cross sections
  of cosmic antiprotons in the light of new data from NA61 and LHCb
  experiments}}, {\emph{ArXiv e-prints} (Feb., 2018) },
  [\href{http://arxiv.org/abs/1802.03030}{{\tt 1802.03030}}].

\bibitem{1997ApJ...490..493N}
J.~F. {Navarro}, C.~S. {Frenk} and S.~D.~M. {White}, \emph{{A Universal Density
  Profile from Hierarchical Clustering}},
  \href{http://dx.doi.org/10.1086/304888}{\emph{\apj} {\bf 490} (Dec., 1997)
  493}, [\href{http://arxiv.org/abs/astro-ph/9611107}{{\tt astro-ph/9611107}}].

\bibitem{2016MNRAS.463.2623H}
Y.~{Huang}, X.-W. {Liu}, H.-B. {Yuan}, M.-S. {Xiang}, H.-W. {Zhang}, B.-Q.
  {Chen} et~al., \emph{{The Milky Way's rotation curve out to 100 kpc and its
  constraint on the Galactic mass distribution}},
  \href{http://dx.doi.org/10.1093/mnras/stw2096}{\emph{\mnras} {\bf 463} (Dec.,
  2016) 2623--2639}, [\href{http://arxiv.org/abs/1604.01216}{{\tt
  1604.01216}}].

\bibitem{2017ApJ...834..110A}
A.~{Albert}, B.~{Anderson}, K.~{Bechtol}, A.~{Drlica-Wagner}, M.~{Meyer},
  M.~{S{\'a}nchez-Conde} et~al., \emph{{Searching for Dark Matter Annihilation
  in Recently Discovered Milky Way Satellites with Fermi-Lat}},
  \href{http://dx.doi.org/10.3847/1538-4357/834/2/110}{\emph{\apj} {\bf 834}
  (Jan., 2017) 110}.

\bibitem{2010JCAP...08..004C}
R.~{Catena} and P.~{Ullio}, \emph{{A novel determination of the local dark
  matter density}},
  \href{http://dx.doi.org/10.1088/1475-7516/2010/08/004}{\emph{\jcap} {\bf 8}
  (Aug., 2010) 4}, [\href{http://arxiv.org/abs/0907.0018}{{\tt 0907.0018}}].

\bibitem{2010A&A...523A..83S}
P.~{Salucci}, F.~{Nesti}, G.~{Gentile} and C.~{Frigerio Martins}, \emph{{The
  dark matter density at the Sun's location}},
  \href{http://dx.doi.org/10.1051/0004-6361/201014385}{\emph{\aap} {\bf 523}
  (Nov., 2010) A83}, [\href{http://arxiv.org/abs/1003.3101}{{\tt 1003.3101}}].

\bibitem{2017arXiv171109616E}
C.~{Evoli}, D.~{Gaggero}, A.~{Vittino}, M.~{Di Mauro}, D.~{Grasso} and M.~N.
  {Mazziotta}, \emph{{Cosmic-ray propagation with DRAGON2: II. Nuclear
  interactions with the interstellar gas}}, {\emph{ArXiv e-prints} (Nov., 2017)
  }, [\href{http://arxiv.org/abs/1711.09616}{{\tt 1711.09616}}].

\bibitem{2017PhRvD..96j3005T}
N.~{Tomassetti}, \emph{{Solar and nuclear physics uncertainties in cosmic-ray
  propagation}},
  \href{http://dx.doi.org/10.1103/PhysRevD.96.103005}{\emph{\prd} {\bf 96}
  (Nov., 2017) 103005}, [\href{http://arxiv.org/abs/1707.06917}{{\tt
  1707.06917}}].

\bibitem{1996ApJ...464..507C}
J.~M. {Clem}, D.~P. {Clements}, J.~{Esposito}, P.~{Evenson}, D.~{Huber},
  J.~{L'Heureux} et~al., \emph{{Solar Modulation of Cosmic Electrons}},
  \href{http://dx.doi.org/10.1086/177340}{\emph{\apj} {\bf 464} (June, 1996)
  507}.

\bibitem{2012AdSpR..49.1587D}
S.~{Della Torre}, P.~{Bobik}, M.~J. {Boschini}, C.~{Consolandi}, M.~{Gervasi},
  D.~{Grandi} et~al., \emph{{Effects of solar modulation on the cosmic ray
  positron fraction}},
  \href{http://dx.doi.org/10.1016/j.asr.2012.02.017}{\emph{Advances in Space
  Research} {\bf 49} (June, 2012) 1587--1592}.

\bibitem{2013PhRvL.110h1101M}
L.~{Maccione}, \emph{{Low Energy Cosmic Ray Positron Fraction Explained by
  Charge-Sign Dependent Solar Modulation}},
  \href{http://dx.doi.org/10.1103/PhysRevLett.110.081101}{\emph{\prl} {\bf 110}
  (Feb., 2013) 081101}, [\href{http://arxiv.org/abs/1211.6905}{{\tt
  1211.6905}}].

\bibitem{2014SoPh..289..391P}
M.~S. {Potgieter}, E.~E. {Vos}, M.~{Boezio}, N.~{De Simone}, V.~{Di Felice} and
  V.~{Formato}, \emph{{Modulation of Galactic Protons in the Heliosphere During
  the Unusual Solar Minimum of 2006 to 2009}},
  \href{http://dx.doi.org/10.1007/s11207-013-0324-6}{\emph{Solar Physics} {\bf
  289} (Jan., 2014) 391--406}, [\href{http://arxiv.org/abs/1302.1284}{{\tt
  1302.1284}}].

\bibitem{2016CoPhC.207..386K}
R.~{Kappl}, \emph{{SOLARPROP: Charge-sign dependent solar modulation for
  everyone}}, \href{http://dx.doi.org/10.1016/j.cpc.2016.05.025}{\emph{Computer
  Physics Communications} {\bf 207} (Oct., 2016) 386--399},
  [\href{http://arxiv.org/abs/1511.07875}{{\tt 1511.07875}}].

\end{thebibliography}
\end{document}